# Spin torque driven dynamics of a coupled two layer structure: interplay between conservative and dissipative coupling

M. Romera[1,2,3], B. Lacoste[4], U. Ebels[1,2,3], L. D. Buda-Prejbeanu[1,2,3]

1. Univ. Grenoble Alpes, INAC-SPINTEC, F-38000 Grenoble, France
2. CEA, INAC-SPINTEC, F-38000 Grenoble, France
3. CNRS, SPINTEC, F-38000 Grenoble, France
4. International Iberian Nanotechnology Laboratory, Braga, Portugal

In this manuscript the general concepts of spin wave theory are adapted to the dynamics of a self-polarized system based on two layers coupled via interlayer exchange (conservative coupling) and mutual spin torque (dissipative coupling). An analytical description of the non-linear dynamics is proposed and validated through numerical simulations. In contrast to the single layer model, the phase equation of the coupled system has a contribution coming from the dissipative part of the LLGS equation. It is shown that this is a major contribution to the frequency mandatory to describe well the most basic features of the dynamics of coupled systems. Using the proposed model a specific feature of coupled dynamics is addressed: the redshift to blueshift transition observed in the frequency current dependence of this kind of exchange coupled systems upon increasing the applied field. It is found that the blueshift regime can only occur in a region of field where the two linear eigenmodes contribute equally to the steady state mode (i.e. high mode hybridization). Finally, a general perturbed Hamiltonian equation for the coupled system is proposed.

## I. INTRODUCTION

The discovery of spin-transfer torque [1] opened new possibilities to manipulate the magnetization in spintronic devices. One important application are spin torque oscillators (STO) which are very promising candidates for the next generation of radiofrequency devices [2, 3, 4]. A standard STO contains an in-plane magnetized ferromagnetic layer which acts as a polarizer and an in-plane magnetized ferromagnetic free layer whose magnetization is driven into large angle self-sustained oscillations under application of current. This precession is usually described in the frame of an independent free layer picture, which considers the magnetization of the free layer independent of the polarizer, whose magnetization is kept fixed (Fig. 1a). The basic features of the in-plane precession steady state of the magnetization of such an independent free layer are well known: a critical current that is almost constant with field and a frequency redshift upon increasing current, almost independent of the applied field [5]. However, in real STOs different dynamic coupling mechanisms between the layers may occur, such as the conservative dynamic dipolar or interlayer exchange (RKKY) interactions or dissipative interactions due to mutual spin-transfer torque (STT). Recent studies have shown that these coupling mechanisms play an important role that cannot be neglected to describe well the features observed experimentally in standard STOs [6,7]. Moreover, these coupling mechanisms can be exploited in the interest of the device performances, for instance to reduce the emitted linewidth [6, 8].

Among the different systems, to study the role of coupling mechanisms on the STT driven steady state excitations, the system of interest here is the *self-polarized* two layer synthetic antiferromagnet (SyF) with interlayer exchange interaction and spin transfer torque acting on both layers simultaneously (see Fig. 1b). Recently first experimental [9] and numerical [10] results have been published on the steady state dynamics for such a self-polarized coupled structure showing a range of interesting phenomena specific to the dynamics of coupled systems. The numerical analysis of Ref. 10 has shown the following interesting features: (i) the critical current of steady state excitations of a self-polarized SyF shows a strong dependence on the applied field, Fig. 2 in Ref. [10], in contrast to the single layer behavior; (ii) the STT excitation is optic like and (iii) a transition in the frequency-current dependence from redshift to blueshift is observed upon increasing the applied field. This latter feature has been observed experimentally in a similar RKKY coupled system using an external polarizer [11]. This frequency dependence thus seems to be a typical feature of the coupled dynamics of a two layer coupled system that is driven by STT into steady state.

To describe the self-sustained magnetization oscillations of a single free layer in magneto-resistive devices, there is a widely used theoretical model [12, 13, 14] that adapted general concepts of spin wave theory to the dynamics of STOs in the single layer approximation. The model, called here spin wave model, explains well the basic features of single layer excitations, in particular the frequency redshift upon increasing current. However a theoretical framework to describe in a general way the non-linear magnetization dynamics of coupled layers is still lacking. The reason is probably the complexity of developing such a model involving coupling. From the studies presented here, it is pointed out that the difficulty arises from the truncation of the analytical equations including different conservative and dissipative terms, in order to define simplified equations, i.e. from the decision of which terms are important to describe well the dynamics of the coupled system. A major result of this work is to provide an approach for this truncation for the specific case of a self-polarized two layer system.

In this manuscript, the spin wave model is adapted in order to describe analytically the self-polarized synthetic ferrimagnet (SyF) in the antiparallel alignment (see Figure 1b). An analytical model that provides the non-linear dynamics of such a coupled magnetic layer system is proposed and validated through numerical simulations upon solving the system of coupled Landau-Lifshitz-Gilbert-Slonczewski (LLGS) equations for the magnetizations $\mathbf{m}_1$ and $\mathbf{m}_2$ of the SyF. For the analytical model it is found that in contrast to the single layer model, the phase equation of the coupled system has a contribution coming from the dissipative part of the LLGS equation. It is shown that this is actually a mandatory contribution to describe well the frequency dependence and the most basic features of coupled dynamics. Indeed a non-linear term coming from the dissipative part of the LLGS equation happens to be the most important contribution to the phase equation and it is crucial for the phase locking of the steady state of the coupled system to occur. Additionally, some particularities of the dynamics of the self-polarized system are discussed. For instance, it is found that in order to describe the STT steady state excitations, both linear eigenmodes of the SyF contribute. Furthermore in the steady state the characteristic relation between the phases of the eigenmodes is such that the *sum* of the phases remains constant, and not the phase difference, as it is usually the case in the dynamics of phase locked systems [12]. Secondly, the redshift to blueshift transition observed in the frequency current dependence upon increasing field of the self-polarized system is addressed using the proposed model, and the key required aspects for a frequency blueshift to occur are pointed out. It is found that the transition comes from the competition between conservative and dissipative terms, where the non-trivial dependence of the non-linear dissipative terms on amplitudes and phases play a key role. The blueshift regime occurs in a region of field where the power ratio $p_1/p_2$ has a weak dependence with current. This happens only where the two linear eigenmodes contribute equally to the steady state mode (i.e. high mode hybridization). Finally, a general perturbed Hamiltonian equation is proposed for this coupled system.

The article is structured as follows: Upon solving LLGS section II describes the magnetization dynamics of the self-polarized SyF system as well as the reference system in which the free layer oscillates independently (single layer approximation without interaction). Section III describes the basis transformation used and develops the general (full) analytical model in the new basis. Section IV presents the numerical analysis used to find the phase relation condition in the steady state and to confirm the validity of the simplified model, which is derived and discussed in section V. Section VI addresses the redshift to blueshift transition in the frequency current dependence and the key aspects for the blueshift regime to occur. Section VII summaries the main conclusions and from these presents a general perturbed Hamiltonian equation for the self-polarized coupled system.

## II. MAGNETIZATION DYNAMICS OF THE SyF

### II.A Coupled LLGS equations of the magnetization

We consider the in-plane-precession (IPP) mode of a self-polarized synthetic ferrimagnet: two in-plane magnetized layers coupled via RKKY exchange interaction and mutual STT (Fig. 1b). The dipolar coupling can also be taken into account, but its effect is smaller than the RKKY coupling for the parameters and dimensions considered here. Therefore, for simplicity it will be neglected in this description.

The magnetization dynamics of the coupled system including the damping-like spin transfer torque is described by the LLG equation enhanced by a term due to the spin transfer torque. For the numerical integration the form of the Landau-Lifshitz (LL) equation is more suitable. After transformation the equations used by this macrospin model are the following:

$$\frac{d\mathbf{m}_1}{dt} = -\gamma'_0 (\mathbf{m}_1 \times \mathbf{H}_{\text{eff},1}) - \gamma'_0 H_{RKKY,1} (\mathbf{m}_1 \times \mathbf{m}_2) - \alpha\gamma'_0 [\mathbf{m}_1 \times (\mathbf{m}_1 \times \mathbf{H}_{\text{eff},1})]$$
$$- \alpha\gamma'_0 H_{RKKY,1} [\mathbf{m}_1 \times (\mathbf{m}_1 \times \mathbf{m}_2)] + \gamma'_0 a_{J,1} [\mathbf{m}_1 \times (\mathbf{m}_1 \times \mathbf{m}_2)] - \alpha\gamma'_0 a_{J,1} (\mathbf{m}_1 \times \mathbf{m}_2)$$
(1a)

$$\frac{d\mathbf{m}_2}{dt} = -\gamma'_0 (\mathbf{m}_2 \times \mathbf{H}_{\text{eff},2}) - \gamma'_0 H_{RKKY,2} (\mathbf{m}_2 \times \mathbf{m}_1) - \alpha\gamma'_0 [\mathbf{m}_2 \times (\mathbf{m}_2 \times \mathbf{H}_{\text{eff},2})]$$
$$- \alpha\gamma'_0 H_{RKKY,2} [\mathbf{m}_2 \times (\mathbf{m}_2 \times \mathbf{m}_1)] - \gamma'_0 a_{J,2} [\mathbf{m}_2 \times (\mathbf{m}_2 \times \mathbf{m}_1)] + \alpha\gamma'_0 a_{J,2} (\mathbf{m}_2 \times \mathbf{m}_1)$$
(1b)

On the right side of equations 1a,b the first two terms are conservative terms, and the last four terms are dissipative terms. Note that the presence of the pseudo-field-like term (last terms in eq. 1a,b) is coming from the transformation of the LLGS equation into the LLS equation. It should not be confused with the field-like torque term ($\sim b_J$) that arises from the out of plane spin torque [15]. $M_{Si}$ and $\mathbf{m}_i$ are the saturation magnetization and the corresponding normalized magnetization vector of the $i$th layer, $\mathbf{H}_{\text{eff},i} = \mathbf{H}_{\text{anis},i} + \mathbf{H}_{\text{dem},i} + \mathbf{H}_{\text{app}}$ is the self-effective field and contains anisotropy, demagnetization and external applied field. Only applied field values are considered that conserve the antiparallel alignment of the SyF. $\mathbf{H}_{\text{RKKY},i} = \frac{J_{\text{RKKY}}}{\mu_0 M_{Si} t_i}$ is the contribution to the $i$th layer given by the RKKY coupling with the other magnetic layer, $\gamma'_0 = \frac{\mu_0 \gamma}{1+\alpha^2}$, $\gamma$ is the gyromagnetic ratio, $\mu_0$ the vacuum permeability, $\alpha$ is the Gilbert damping constant and $a_{J,i} \sim \frac{j_{app}}{M_{Si} t_i}$ is the STT prefactor and is proportional to the applied current density ($j_{\text{app}}$). Note that the signs of the terms of the spin torque terms (last two terms in eq. 1a,b) are opposite in eq. 1a and eq. 1b. This is due to the mutual STT effect (direct electrons and back-scattered respectively). No further external spin transfer torque from a polarizer is considered.

The equations 1a,b are solved numerically for two cylindrical layers with 85nm in diameter, separated by a non-magnetic spacer layer. The corresponding material parameters considered are: saturation magnetization $M_{S1}=M_{S2}=1000$ kA/m, uniaxial anisotropy constant along the x-axis $K_{u1}=K_{u2}=1000$ J/m$^3$, thickness $t_1=3.9$nm, $t_2=3.95$nm, damping constant $\alpha_1=\alpha_2=0.02$ and interlayer exchange energy $J_{\text{RKKY}}=-1\cdot 10^{-4}$ J/m$^2$. The parameters have been used in our previous numerical studies [10] to obtain the trajectories, the state diagrams and the frequency current dependence of a self-polarized SyF. From these previous studies, the value of $J_{\text{RKKY}}$ has been chosen here so that it gives rise to a wide range of steady state excitations (IPP) but avoids the chaotic behavior for larger current densities.

**II.B Numerical results**

In order to model the behavior of a single layer, the magnetization dynamics of $\mathbf{m}_1$ is blocked and both RKKY coupling and mutual spin torque are suppressed in the simulations (only spin transfer torque on layer 2 is considered). In this way the well-known single layer picture is recovered, where $\mathbf{m}_1$ behaves as an external polarizer with fixed magnetization orientation and $\mathbf{m}_2$ can be considered as a single free layer (Fig. 1a). As expected, the frequency-current behavior of this system shows a typical redshift behavior at all fields (Fig. 1c).

The numerical results of the coupled SyF system of Fig. 1b, solving equations 1a and 1b, are given in Fig. 1d. As already shown numerically in ref 10, there is a transition in the frequency-current dependence from redshift to blueshift upon increasing the applied field. This type of transition has been observed experimentally [11] in RKKY coupled layers with external polarizer.

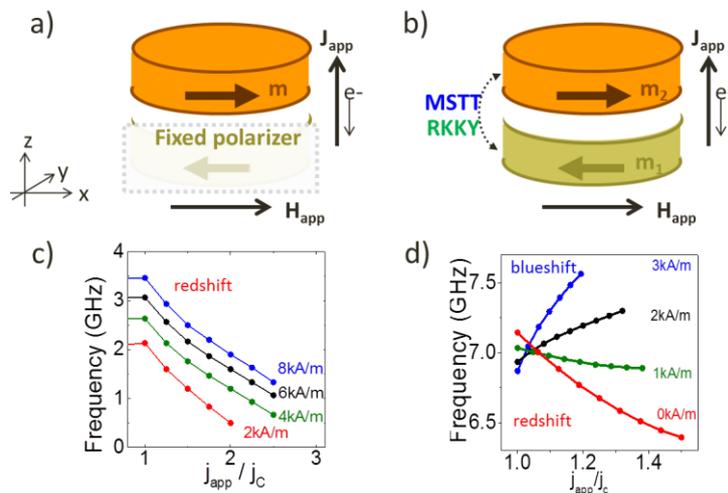

Figure 1: Schematics of (a) the single layer picture and (b) the self-polarized SyF system. Frequency versus supercriticality at T=0 K at different values of field for (c) the single layer case and (d) for the self-polarized SyF obtained by solving numerically the LLGS equation.

### III. TRANSFORMED COUPLED EQUATIONS

In order to understand the origin of the difference between the frequency-current dependencies of a single layer and SyF structure we now turn to deriving analytical equations for the power and phases. To describe the magnetization dynamics under spin transfer torque of a single free layer, the spin wave theory [13] was adapted [12, 14], upon transforming the LLGS equation from the magnetization variables (m$_x$,m$_y$,m$_z$) to a new basis of complex $e$variables. In this basis, the magnetization trajectories are described by normalized circles. A generalization of this model for the coupled system is proposed here. It includes several canonical transformations which will be applied to the conservative and the dissipative parts separately. The details of the proposed transformation are presented elsewhere [16], while here only the main steps of the analytical development are shown and the focus is on (1) determining the leading terms of the analytical development by comparison with the numerical simulations, (2) defining a truncated simplified model with only the terms required to describe well the dynamics of the coupled system and (3) understanding the origin of the redshift to blueshift transition in the frequency-current dependence.

**III.A Conservative terms**

The first transformation applied is similar to the transformation used in the single layer model [12, 17]. It connects the cartesian magnetization components $m_{x,y,z}(t)$ to the complex variables $a(t)$ and $a^*(t)$, a new basis where the single layer magnetization dynamics can be described in 2 dimensions (see appendix 1). The Hamiltonian $\mathcal{H}$ obtained can be expressed separating terms that correspond to layer 1, terms that correspond to layer 2 and RKKY coupling terms (first row in eq. 2). Alternatively it can be expressed separating the quadratic part of the Hamiltonian $\mathcal{H}_2$, from the fourth order terms ($\mathcal{H}_4$) and the higher order terms (second row in eq. 2):

$$\mathcal{H}(a_1,a_1^*,a_2,a_2^*) = \mathcal{H}_{m1}(a_1,a_1^*) + \mathcal{H}_{m2}(a_2,a_2^*) + \mathcal{H}_{RKKY}(a_1,a_1^*,a_2,a_2^*) \qquad (2)$$
$$= \mathcal{H}_2(a_1,a_1^*,a_2,a_2^*) + \mathcal{H}_4(a_1,a_1^*,a_2,a_2^*) + O(6)$$

Then, a linear transformation $T_{ab}$, with $\mathbf{a}=T_{ab}\cdot\mathbf{b}$ is applied [16] to diagonalize the quadratic part of the Hamiltonian $\mathcal{H}_2$ to the form:

$$\mathcal{H}_2(b_1,b_1^*,b_2,b_2^*) = \omega_1|b_1|^2 + \omega_2|b_2|^2 \qquad (3)$$

where $b_1$ and $b_2$ form the new basis corresponding to the linear eigenmodes of the coupled system. While $a_1$ is related exclusively to $\mathbf{m_1}$, and $a_2$ to $\mathbf{m_2}$, the new eigenmodes $b_1$ and $b_2$ involve both layers $\mathbf{m_1}$ and $\mathbf{m_2}$, which oscillate simultaneously. $b_1$ and $b_2$ correspond to the optical and acoustic modes respectively in the linear regime of a synthetic antiferromagnet in the antiparallel alignment.

The transformation $T_{ab}$ is then applied to the full Hamiltonian. In the general model considered in this section, all terms are taken into account. The full Hamiltonian reads:

$$\mathcal{H} = \omega_1|b_1|^2 + \omega_2|b_2|^2 + \frac{1}{2}N_1|b_1|^4 + \frac{1}{2}N_2|b_2|^4 + T|b_1|^2|b_2|^2 \qquad (4)$$
$$+ \frac{1}{2}\left(Fb_1b_1b_2b_2 + Sb_1b_1b_2^*b_2^* + c.c.\right) + H_{4r} + H_{hr}$$

The coefficients, $\omega_1$, $\omega_2$, $N_1$, $N_2$, $T$, $F$, $S$, are real and depend on the material parameters related to $\mathbf{H}_{eff,i}$ and to the RKKY coupling $\mathbf{H}_{RKKY,i}$, but are independent of the current and of the Gilbert damping constant.

The first two terms ~ $\omega_1$, $\omega_2$ give rise to the linear dynamics and the terms ~$N_1$, $N_2$, $T$ give rise to non-linear frequency shifts. From the fourth order terms, the F and S terms have been distinguished. They are called parametric interaction between modes (they allow energy transfer between modes) in analogy to a previous work about multimode generation regimes in single layer excitations [17].

From this Hamiltonian the conservative (C) part of the equation of motion is obtained:

$$\dot{b}_i\bigg|_C = -j\frac{\partial\mathcal{H}}{\partial b_i^*} = -j(\Omega_i + \Psi_i)b_i \qquad (5)$$

using the following definitions:

$$\Omega_1 = \omega_1 + N_1 p_1 + T p_2$$
$$\Psi_1 = F\frac{b_1^* b_2^{*2}}{b_1} + S\frac{b_1^* b_2^2}{b_1} + \frac{1}{b_1}\frac{\partial(H_{4r}+H_{hr})}{\partial b_1^*}$$
$$\Omega_2 = \omega_2 + N_2 p_2 + T p_1$$
$$\Psi_2 = F\frac{b_2^* b_1^{*2}}{b_2} + S\frac{b_2^* b_1^2}{b_2} + \frac{1}{b_2}\frac{\partial(H_{4r}+H_{hr})}{\partial b_2^*}$$

where the powers $p_i$ are defined as $p_i=|b_i|^2$, $\Omega_{1,2}$ are real and $\Psi$ is complex through its dependence on $b_1$, $b_2$, $b_1^*$ and $b_2^*$ while F and S are real numbers.

### III.B Dissipative terms

The same transformation $T_{ab}$ is applied to the dissipative part of the LLGS equation, giving rise to the dissipative (D) part of the equation of motion of $b_1$, $b_1^*$, $b_2$, $b_2^*$. For $b_1$ it can be described in a general way:

$$\begin{aligned} \dot{b}_1\Big|_D &= \sum_{k,l,m,n} f_{klmn,1} b_1^k b_2^l b_1^{*m} b_2^{*n} \\ &= -(j\Sigma_1 + \Gamma_1 + \Pi_1) b_1 \end{aligned} \quad (6)$$

Here $\Sigma_i$ is a complex function that arises from the contribution of the pseudo-field like term, $\Gamma_i$ is a complex function that includes the dissipative diagonal terms and $\Pi_i$ is a complex function that includes the dissipative off-diagonal terms:

$$\Pi_1 = \frac{1}{b_1}(\Xi_1 b_1^* + \Theta_1 b_2 + \Delta_1 b_2^*)$$

$$\Pi_2 = \frac{1}{b_2}(\Xi_2 b_2^* + \Theta_2 b_1 + \Delta_2 b_1^*)$$

The coefficients $\Gamma_i$, $\Xi_i$, $\Theta_i$, $\Delta_i$ are complex functions depending on material parameters, injected current value, the damping constant, higher order terms of the amplitude $|b_k|^n$ and phases $\phi_j$, with $j=1,2$. Similar equations are derived for $b_1^*$, $b_2$, $b_2^*$.

### III.C Full transformed equations

With this, the general perturbed Hamiltonian equation of the self-polarized SyF system reads as:

$$\begin{aligned} \dot{b}_1 &= -[j(\Omega_1 + \Psi_1 + \Sigma_1) + \Gamma_1 + \Pi_1] b_1 \\ \dot{b}_2 &= -[j(\Omega_2 + \Psi_2 + \Sigma_2) + \Gamma_2 + \Pi_2] b_2 \end{aligned} \quad (7)$$

Considering solutions of the form $b_i(t) = |b_i(t)| \exp[-j\phi_i(t)]$ with $i=1,2$, two equations for the amplitudes and two equations for the phases are obtained:

$$\begin{cases} \dot{p}_1 = -2[\mathrm{Re}(\Gamma_1) + \mathrm{Im}(\Psi_1) + \mathrm{Im}(\Sigma_1) + \mathrm{Re}(\Pi_1)] p_1 \\ \dot{p}_2 = -2[\mathrm{Re}(\Gamma_2) + \mathrm{Im}(\Psi_2) + \mathrm{Im}(\Sigma_2) + \mathrm{Re}(\Pi_2)] p_2 \end{cases} \quad (8)$$

$$\begin{cases} \dot{\phi}_1 = \Omega_1 + \mathrm{Re}(\Psi_1) + \mathrm{Re}(\Sigma_1) + \mathrm{Im}(\Gamma_1) + \mathrm{Im}(\Pi_1) \\ \dot{\phi}_2 = \Omega_2 + \mathrm{Re}(\Psi_2) + \mathrm{Re}(\Sigma_2) + \mathrm{Im}(\Gamma_2) + \mathrm{Im}(\Pi_2) \end{cases} \quad (9)$$

### III.D Discussion

From equations 8 and 9 we can recover the well-known equations of a single layer upon setting the coupling between layers to zero (which results in $\Psi$, $\Sigma$, $\Pi=0$):

$$\frac{\mathrm{d}p}{\mathrm{d}t} = -2\Gamma p \quad (10a)$$

$$\frac{\mathrm{d}\phi}{\mathrm{d}t} = \omega_0 + Np \quad (10b)$$

Here $\Gamma$ corresponds to the sum of damping and spin torque dissipative contributions $\Gamma = \Gamma_+ + \Gamma_-$ [12]. Notice that no terms are discarded from equations 8 and 9, unlike the model for a single layer (eq. 10), for which only resonant terms (terms whose contribution averaged over one oscillation period is non-zero) are kept after truncation.

There are two important differences between the equations 8 and 9 describing the SyF and the simplified eq. 10 describing a single layer: firstly, in the phase equation 9, conservative but also dissipative terms appear, unlike eq. 10b, which contains only conservative terms; correspondingly, in the power equation 8, also conservative terms appear, unlike eq. 10a, which contains only dissipative terms. Secondly, both phase and power equations (eq. 8 and 9) depend explicitly not only on the power, but also on the phases (for example there will be a term proportional to sin($\phi_1+\phi_2$) within Im($\Pi_i$)), unlike the simplified model for a single layer (eq. 10a and 10b).

Furthermore, while $b_1$ and $b_2$ are respectively the optic and acoustic eigenmodes in the linear regime, in the non-linear regime they are no more eigenmodes. This can be seen in equation 7: the d$b_i$/dt equations depend on both $b_1$ and $b_2$. This means that the steady state mode will be a combination of $b_1$ and $b_2$ i.e. the final steady state mode will never be a pure $b_1$ (optic) mode or a pure $b_2$ (acoustic) mode but it will be a mix of both (hybridization). This confirms our previous findings from numerical studies in Ref. [10] of the self-polarized SyF that revealed that the steady state mode could not be described by a pure optic mode for which the $m_y$ ($m_z$) components of the two magnetic layers would be in-phase (out-of-phase). In section VI, it will be shown that the weight of the contribution of each mode to the final mixed mode for a given field depends on the applied current.

In order to relate experimentally or numerically observed features to physical parameters, a simplified and more manageable model is desirable, i.e. equations 8, 9 should be truncated. This is not an easy task, because there are no evident criteria to decide which terms dominate. In the next sections IV and V, we present an approach to determine which terms have a significant impact on the dynamics. This approach consists on taking advantage of numerical simulations to (i) find selection criteria regarding specific phase relations and power relations fulfilled in the steady state regime and (ii) compare the value of each analytical term with the total frequency value obtained numerically, in order to evaluate the importance of each term.

## IV. NUMERICAL EVALUATION OF PHASE AND POWER IN THE STEADY STATE

In order to find conditions for the phase and amplitude in the steady state we proceed as follows. First, we solve numerically the coupled LLGS equation Eq. 1, which provides the time evolution of the six magnetization components $m^i_x(t)$, $m^i_y(t)$, $m^i_z(t)$ with i=1,2 for both layers $\mathbf{m}_1$ and $\mathbf{m}_2$. Then we apply to the solutions $\mathbf{m}_1(t)$ and $\mathbf{m}_2(t)$ the two transformations developed in section III (**m**-variables into a-variables and $T_{ab}$) to obtain the two complex variables $b_1(t)$ and $b_2(t)$ from which the respective powers $p_1(t)$ and $p_2(t)$ and phases $\phi_1(t)$ and $\phi_2(t)$ are obtained. This evaluation of $b_1(t)$ and $b_2(t)$ corresponds to considering all terms in eq. 8 and eq. 9. The numerical analysis focuses on the range of positive fields up to *H*=3kA/m and positive current densities, conditions for which the redshift to blueshift transition is observed and for which the SyF remains in its antiparallel alignment.

### IV.A Trajectories

In Fig 2a, the in-plane projection of $\mathbf{m}_1(t)$ and $\mathbf{m}_2(t)$ onto the y-z plane is shown, for field and current conditions (*H*=0 A/m and $j_{app}$=450 x $10^{10}$ A/m$^2$) that lead to steady state in the redshift regime. Both layers are excited at almost equal amplitude, but precess such that their relative phases are close to an optic like mode (see Fig. 3 in Ref. 10). In Fig. 2b, the corresponding variables $b_1(t)$ and $b_2(t)$ after transformation are plotted in the complex plane. For all current and field conditions explored numerically, it is found that the steady state is dominated by mode $b_1$ (which corresponds to the optic mode in the linear regime), i.e. the power $p_1$ is much larger than the power $p_2$, but both $p_1$ and $p_2$ are non-zero. This confirms the conclusion from eq. 7 to eq. 9 that in the steady state both modes contribute.

### IV.B Phase relation

From Fig. 2b it can also be seen that the trajectory of $b_1$ is almost a circle while this is not the case for $b_2$. But more importantly, the rotation sense of $b_2$ is opposite to the one of $b_1$. This is confirmed when regarding the time evolution of the numerically calculated phases $\phi_1(t)$ and $\phi_2(t)$ as shown in Fig. 2c. It can be seen that in the transient regime (t≤6ns) the phases $\phi_1$ and $\phi_2$ have the same sign. In contrast, in the steady state (t>6ns) $\phi_1$ and $\phi_2$ are coupled so that their time-derivatives have equal absolute value but opposite sign and the average of their sum is constant. These two conditions are fulfilled for the whole steady state range and can be summarized by equations 11 and 12:

$$\left|\frac{\mathrm{d}\phi_1(t)}{\mathrm{d}t}\right| = \left|\frac{\mathrm{d}\phi_2(t)}{\mathrm{d}t}\right| \qquad (11)$$

$$\langle \phi_1(t)+\phi_2(t) \rangle = \langle |\phi_1(t)|-|\phi_2(t)| \rangle = \langle \psi \rangle = \mathrm{const} \qquad (12)$$

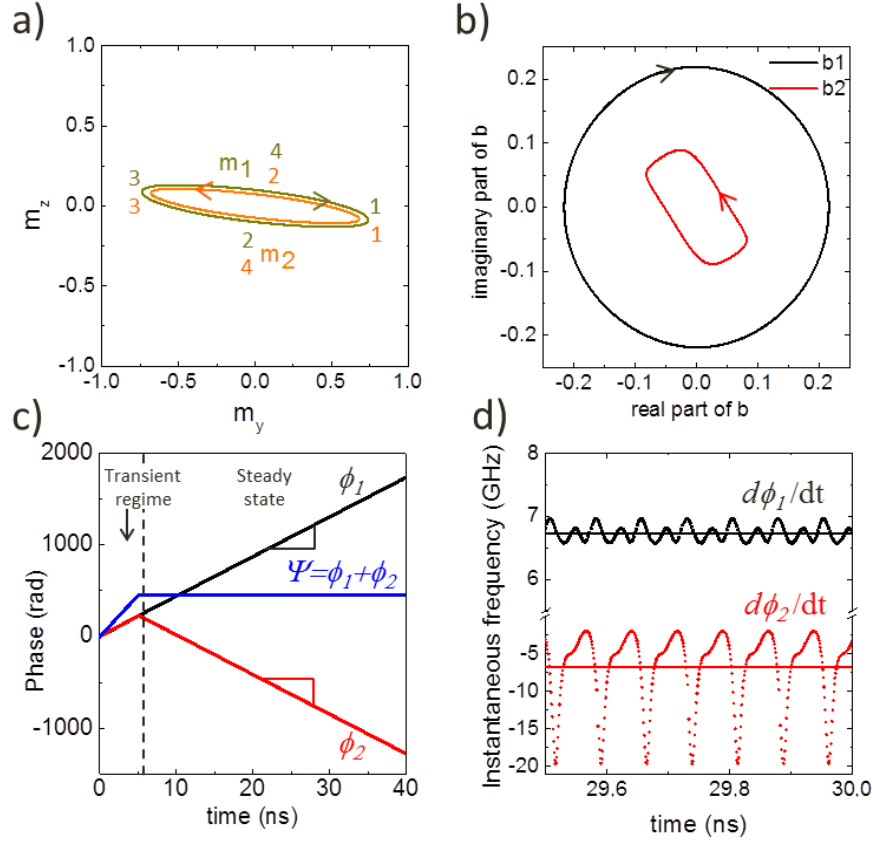

Figure 2: Analysis of the steady state dynamics in (a) the $m_{x,y,z}$ components and (b-d) the $b$-basis, obtained by numerical simulations for zero applied field and $j_{app}$= 450 x $10^{10}$ A/m$^2$. (b) Imaginary vs real part of $b_1$ and $b_2$. Time evolution of the phases (c) and instantaneous frequencies (d). Horizontal lines in (d) correspond to the average frequency values.

Equation 11 states that the absolute value of the average frequency is the same for variables $b_1$ and $b_2$. In order to give an intuitive interpretation of this condition let us go one step back to the original basis $\mathbf{m}_1$ and $\mathbf{m}_2$. It is clear that when the coupled system is in steady state both layers $\mathbf{m}_1$ and $\mathbf{m}_2$ oscillate at the same frequency due to the coupling: if the current polarity is such that layer 1 is excited, layer 2 would follow due to the RKKY coupling and therefore at steady state both oscillate together and at the same average frequency. Equation 11 is basically the analog picture in the basis $b_1$ and $b_2$, where now $b_1$ and $b_2$ are coupled through non-linear and higher order resonant terms from eq. 8 and 9. It is noted that large variations of the instantaneous frequency $\dot{\phi}_2(t)$ occur in comparison to $\dot{\phi}_1(t)$ (Fig. 2d). We relate this feature to the fact that the steady state is dominated by $b_1$ and it is mostly $b_2$ which adapts its phase (frequency), as can be seen in Fig. 2c

It is pointed out here, that equation 12 has been found to be a specific condition of the steady state of the self-polarized SyF system, and it does not apply to all coupled systems (in contrast to eq. 11, which still applies). For a coupled two layer system with external polarizer [16] the same analysis shows that the phase relation is such that the phase difference is constant:

$$\langle \phi_1(t) - \phi_2(t) \rangle = \langle \psi_{ex} \rangle = \text{const} \qquad (13)$$

From this it is concluded that one has to investigate the phase relationship separately for each situation. Although eq. 13 is a quite general phase relation in the dynamics of coupled or synchronized systems, it is the self-polarized system studied here which seems to be a particular case fulfilling eq. 12. This is a major finding of this work.

In the following it will be shown that this particular phase relation in the steady state regime is linked to the fact that the main contribution for this system are the dissipative off-diagonal resonant terms ($\Pi_i$ in eq. 7 and correspondingly Im($\Pi_i$) in eq. 9 ).

**IV.C Range of the phase relation ⟨ψ⟩**

From the numerical simulations it was found that the average value of $\psi$ over one period ⟨ψ⟩, takes only a specific range of values. That is from almost $2\pi/3$ $(+2n\cdot\pi)$ at the critical current, towards $\pi/2$ $(+2n\cdot\pi)$ at the onset of instability and chaos that is found at higher currents. Fig. 3 represents this range as a function of supercriticality $j_{app}/j_C$, for different values of applied field. Here $j_{app}$ is the applied current density and $j_C$ the critical current density. It can be seen that the current dependencies of ⟨ψ⟩ for different field values superpose, i.e. they take the same values in the redshift H=0, 1kA/m) as well as in the blueshift regimes (H=3kA/m).

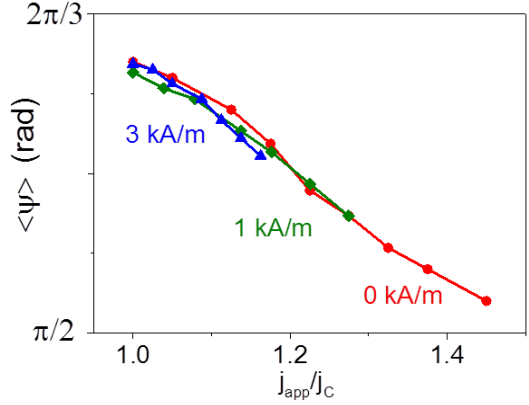

Figure 3: ⟨ψ⟩ evolution with supercriticality at different fields.

Upon increasing current at a constant field ⟨ψ⟩ decreases towards $\pi/2$. As it will be shown in section VI, this corresponds to a stronger mode hybridization (high mode mixing).

In the next section the conditions of eq. 11 and 12 will be used to truncate eq. 9, i.e. to simplify eq. 9 as much as possible, keeping only the terms which are necessary to describe properly the coupled dynamics. Before doing so and to conclude this section, it is worth mentioning a further observation from the analysis of the numerical solutions.

**IV.D Damped mode**

The numerical and analytical analysis above has considered only the STT driven steady state excitation. The numerical analysis provides also access to the damped mode that coexists with the steady state mode, by introducing thermal fluctuations. This damped mode is at much lower frequency than the STT mode. The corresponding frequencies as a function of supercriticality ($j_{app}/j_C$) are shown in Fig. 4 for different values of applied field. Interestingly, a strong frequency redshift is observed for the lower frequency damped mode upon increasing current. This redshift increases upon increasing the applied field and a reduction of values as high as 3 GHz is observed. The redshift is an indication that a strong mode mixing of $b_1$ and $b_2$ exists not only for the steady state mode but that also that the damped mode is affected.

The reason for such a strong frequency shift can be interpreted as follows: due to the large amplitude of the spin torque driven excitation, the damped mode (which is at a different frequency than the steady state mode) sees modified averaged internal fields, such as a modified average RKKY exchange field. These modified fields give rise to the frequency shift. This again can be understood better when going a step back to the original basis $\mathbf{m}_1$ and $\mathbf{m}_2$. If $\mathbf{m}_1$ oscillates at large amplitudes at $\omega_1$, then for instance the effective field seen by the magnetization of the damped mode $\mathbf{m}_2$ which has a frequency $\omega_2$ would be different as compared to the situation of $\mathbf{m}_1$ being static.

In summary, this result shows quite generally that the frequencies of the damped modes of a system can be strongly affected, when one mode is driven into large angle steady state excitations. This change of frequency does not occur only for when frequencies cross as discussed in ref [6, 18].

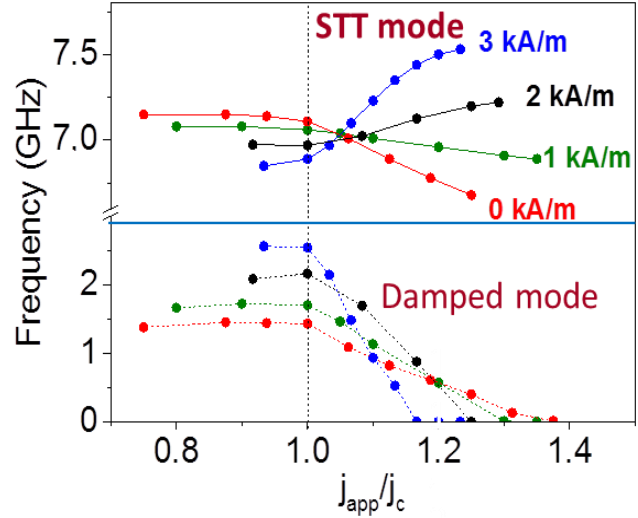

Figure 4: Frequency of the two modes of the coupled system (the higher frequency steady state mode and the damped lower frequency mode) as a function of the supercriticality at different fields.

## V. SIMPLIFIED MODEL

The numerical analysis of $b_1(t)$ and $b_2(t)$ from section IV, corresponds to considering all terms of eq. 9. The next step consists now in identifying the terms in eq. 9 which integrated over one oscillation period are non-zero (resonant terms) with a significant impact on the steady state dynamics of the coupled system.

For this we insert the conditions of eq. 11 and eq. 12, into eq. 9. Here it is clear that terms depending explicitly on $\psi$ must be included in the truncated model. In addition, coupling between modes gives rise to non-linear terms with phase relations different than $\psi$ which may become resonant due to non-obvious relations of phases and powers.

The prefactors within eq. 9 are defined via the variables transformation and are calculated for the same material parameters, current and field as used for the full numerical solutions. Then, using the values of $b_i$ from the numerical simulations, each term of eq. 9 is calculated independently and averaged over one period. In this way, their relative contributions to the average frequency $\langle d\phi/dt \rangle$ can be evaluated for different currents and fields. This allows us to identify the terms playing a dominant role in the dynamics and to remove the negligible terms to define a simplified model. The requirement is that the dominant terms describe the frequency values as well as the redshift to blueshift transition observed in the frequency current dependence upon increasing field, Fig. 1d, which is a non-linear effect specific of coupled dynamics.

**V.A Truncated Equation**

From this analysis, which is described in detail in section V.B, we obtain the following truncated equation for the phases:

$$\begin{cases} \dot{\phi}_1 = \Omega_1 + Fp_2 \cos(2\psi) + \Lambda_1 + \sqrt{\dfrac{p_2}{p_1}} \left[ \vartheta_1 \cos(\psi) + \widetilde{\Theta}_1 \sin(\phi_1 - \phi_2) + \widetilde{\Delta}_1 \sin(\psi) \right] \\ \dot{\phi}_2 = \Omega_2 + Fp_1 \cos(2\psi) + \Lambda_2 + \sqrt{\dfrac{p_1}{p_2}} \left[ \vartheta_2 \cos(\psi) + \widetilde{\Theta}_2 \sin(\phi_1 - \phi_2) + \widetilde{\Delta}_2 \sin(\psi) \right] \end{cases} \qquad (14)$$

The first three terms of eq. 14 are conservative terms: the term $\Omega_i$ contains the linear dynamics and the non-linear frequency shifts (first row in eq. 4); the second term is coming via $\mathrm{Re}(\Psi_i)$ from the *F*-term of the second row of eq.4; the term $\Lambda_i = U_i p_i^2 + V_i p_j^2 + W_i p_i^2 p_j^2$, with $i,j=1,2$ and $i \neq j$, arises via $\mathrm{Re}(\Psi_i)$ from higher order terms of the Hamiltonian ($\mathcal{H}_\mathrm{hr}$) and it is a contribution due to the RKKY coupling. It is noted that the $U_i$, $V_i$ and $W_i$ coefficients are usually larger than $N_i$, which explains why these terms have a certain impact despite being of higher order. Also, $V_i$ and $W_i$ are generally larger than $U_i$, so they are especially important when there is a considerable mode mixing, as it is here the case. The fact that higher order terms are necessary to describe the frequencies evidences the complexity of the dynamics of the coupled system.

The last three terms of eq. 14 are dissipative terms, the term $\sim \vartheta_i$ is coming from the pseudo field-like torque ($\mathrm{Re}(\Sigma_1)$ in eq. 9) and the two terms $\sim \widetilde{\Theta}_i, \widetilde{\Delta}_i$ are coming from the off diagonal terms ($\mathrm{Im}(\Pi_i)$ in eq. 9). The terms $\sim \widetilde{\Theta}_i, \widetilde{\Delta}_i$ correspond to the truncation of $\Theta_i, \Delta_i$ respectively (i.e. the resonant terms with a significant impact in the dynamics). The coefficients $\vartheta_i, \widetilde{\Theta}_i, \widetilde{\Delta}_i$ are real functions of the form $\chi_i = \chi_{i,0} + \chi_{i,1}|b_1|^2 + \chi_{i,2}|b_2|^2$.

Note that in the truncated eq. 14 there is no conservative term related to the S-term of the second row of eq. 4. It is mentioned that such a term is an important conservative term in most coupled systems. Interestingly, it can be considered as negligible in the particular case of the self-polarized SyF discussed here. This is a direct consequence of the fact that the phase condition for dynamics is given by eq. 12 instead of by the more common phase relation eq. 13. The term that is derived from the S-term of the Hamiltonian depends on the phase difference. These terms are in principle non-resonant and can be neglected with respect to the other terms. This has been further confirmed by evaluating its relative contributions to the average frequency, as it will be shown later on in Fig. 5, section V.B. On the other hand, the term that comes from the F-term of the Hamiltonian, depends on the sum of the phases and as it will be shown in the following, it is very important for the self-polarized system.

**V.B Comparison to numerics**

After averaging the different terms from the right-hand-side of the instantaneous frequencies $d\phi/dt$ of eq. 9, the simplified model eq. 14 was obtained by evaluating the impact of the different terms of eq. 9 on the values of the frequency and on the frequency-current dependence at two different fields as will now be discussed. The relative weight of each term with respect to the frequency calculated from the untruncated numerical solution of eq. 1 then provided the criteria to keep or neglect a term. These untruncated numerical solutions are given by the black lines in Fig. 5a-d.

**Redshift range:** We will begin by analyzing the low field range (Figs. 5(a-b)), where the numerical analysis shows a typical frequency redshift behavior upon increasing current (black symbols in Figs. 5(a-b)). Let us start by analyzing the $<d\phi_1/dt>$ equation (Fig. 5a). Green symbols in Fig. 5a show the values given only by the first conservative term ($\sim \Omega_1$) of the $<d\phi_1/dt>$ equation, which contains the non-linear frequency shifts. From Fig. 5a $\Omega_1$ may seem to be enough to describe reasonably well the frequency values and the redshift behavior. However, from Fig. 2c and eq. 11 it is known that the evaluation of $<d\phi_2/dt>$ must give the same frequency values. Fig. 5b clearly shows that the equivalent term that contains the linear and non-linear frequency shifts ($\Omega_2$ – green symbols in Fig.5b) is not enough, and that additional terms must be considered. A further analysis shows that the main contribution to the $<d\phi_2/dt>$ equation is the term $\sim \widetilde{\Delta}_2$. It is the largest contribution to the frequency values (see violet symbols in Fig. 5b) and it is the crucial term to lock the phases so that eq. 11 and eq. 12 are fulfilled. Interestingly, this major term is coming from the dissipative part of the LLGS equation. The main reason behind its dominant contribution is the power ratio prefactor $p_1/p_2$, which is very large when the steady state mode is dominated by $b_1$. In order to approach the numerical values more terms must be included. The orange symbols in Fig. 5b show the values of $<d\phi_2/dt>$ obtained analytically including all the terms in equation 14.

**Blueshift range:** Let us analyze now the high field range (Figs. 5(c-d)), where the numerical analysis shows a blueshift behavior (black symbols in Figs. 5(c-d)). Again, we first focus on $<d\phi_1/dt>$ (Fig. 5c). Green symbols in Fig.5c show the values obtained considering only the term $\Omega_1$ which contains the non-linear frequency shifts. In contrast to the redshift range, it can be clearly seen that the $\Omega_1$ term does not describe well the behavior of $<d\phi_1/dt>$ at high fields. In particular, it only gives rise to a frequency redshift. To describe the blueshift it is necessary to add the terms $\sim \widetilde{\Theta}_i, \widetilde{\Delta}_i$ of eq. 14 (see violet symbols in Fig. 5c), which are both off-diagonal terms coming from the dissipative part of the LLGS equation (Im($\Pi_1$) in eq. 9). The most important correction to $\Omega_1$ in the analysis of <d$\phi_1$/dt> is the term ~ $\widetilde{\Delta}_1$ which is the same dissipative term that we just highlighted in the analysis of <d$\phi_2$/dt> at low fields. Similarly, the analysis of <d$\phi_2$/dt> at high fields shows that the $\Omega_2$ term (green symbols in Fig. 5d) only yields a redshift, while the term ~ $\widetilde{\Delta}_2$ must be included to describe the blueshift behavior (violet symbols Fig. 5d). In order to approach the numerical values more terms must be included. The orange symbols in Fig. 5d show the values of <d$\phi_2$/dt> obtained by including all terms in equation 14.

Red symbols in Fig. 5 represent the contribution arising from the S-term of the Hamiltonian. As it can be seen it gives a minor contribution which can be considered negligible, the reason why it is not included in the simplified model.

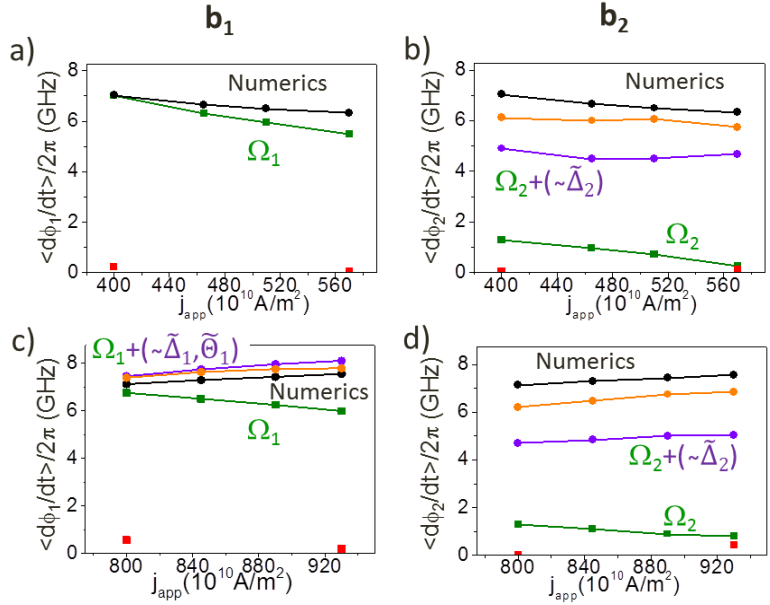

Fig. 5: Frequency current dependence at (a-b) $H$=0kA/m and (c-d) $H$=3kA/m obtained from the (a-c) <d$\phi_1$/dt> and (b-d) <d$\phi_2$/dt> equations. Black symbols correspond to values obtained numerically using eq. 9, and the rest of the symbols correspond to values obtained analytically including different terms in the simplified model of eq. 14 (see text). Orange symbols correspond to the values obtained using the full equation 14. Red symbols represent the weight of the term which results from the S-term of the Hamiltonian, which has not been included in the truncated model.

The role of the different terms of eq. 14 and the reason why they all need to be included can be summarized as follows: at low field and current the terms $\Omega_i + \sqrt{\frac{p_j}{p_i}}[\vartheta_i \cos(\psi) + \widetilde{\Delta}_i \sin(\psi)]$ would be enough to describe well the frequency values for both <d$\phi_1$/dt> and <d$\phi_2$/dt>. If the current is increased at low fields, the terms $Fp_j \cos(2\psi)$ and $\Lambda_i$ become important and necessary for the model to describe a redshift behavior for both <d$\phi_1$/dt> and <d$\phi_2$/dt> equations. At high fields and currents, besides those terms, the term $\sqrt{\frac{p_2}{p_1}}\widetilde{\Theta}_1 \sin(\phi_1 - \phi_2)$ is necessary to describe the blueshift behavior in <d$\phi_1$/dt>.

To conclude this section, it is noted that doing a similar analysis for an externally polarized coupled system, having replaced the condition given in eq. 12 by the corresponding definition of $\psi_{ex}$ (eq. 13), very similar equations to those shown in eq. 14 are obtained, with one interesting and important difference: the term $\propto \sqrt{\frac{p_j}{p_i}}\sin(\psi_{ex})$ of the <d$\phi$/dt> equation of the non-excited mode (<d$\phi_2$/dt> if $b_1$ is dominantly excited) has opposite sign than in eq. 14. This different sign is responsible for the time-derivative of the phases of $b_1$ and $b_2$ to have the same sign in that system (or what is the same, equation 13 being valid instead of equation 12). Interestingly, this term is also the major contribution in the externally polarized system.

## VI. REDSHIFT TO BLUESHIFT TRANSITION: DISCUSSION

For a given field, the redshift or blueshift character of the frequency is given by the current dependence of the different terms involved. Fig. 5 already indicates that it can actually be considered as the result of the relative contributions from conservative and dissipative terms that change as a function of current. The dependence of the different terms on current is mediated through (i) the explicit power dependence, which is the major contribution and (ii) the dissipative term coefficients, which are current dependent.

In order to understand the redshift to blueshift transition upon increasing field one should analyze how the current dependence of these terms change upon increasing field. In other words, one should analyze the change with field of the balance between the terms that give redshift contributions and the terms that give blueshift contributions.

The current dependence of the dissipative coefficients does not change so much with field and therefore it has a minor effect in the discussed feature. In contrast, the current dependence of the power ratio prefactors has been found to change considerably between low and high fields. This will be shown to be the key point to explain the redshift to blueshift transition.

This section is organized as follows: In (A) the *mode asymmetry* is introduced, which quantifies the level of mixing of the steady state mode and will be helpful to understand the current dependence of the powers. The different dependencies at low and high fields will be discussed in (B). Finally in (C) the consequences of these differences on the frequency-current dependence and the reasons behind the redshift to blueshift transition will be discussed.

**VI.A Mode asymmetry and mode mixing**

Let us first introduce the mode asymmetry [19], which is given by

$$m = \frac{p_1 - p_2}{p_1 + p_2} \tag{15}$$

$m \sim 1$ corresponds to a steady state where only eigenmode $b_1$ is excited, $m \sim -1$ corresponds to a steady state where only eigenmode $b_2$ is excited and intermediate values correspond to a hybridized steady state mode. The closer the mode asymmetry is to a value $m \sim 1$ ($m \sim -1$) the larger the contribution of eigenmode $b_1$ ($b_2$) to the hybridized steady state. Fig. 6 shows the dependence of $m$ on $\psi$ along one period of the trajectory for different currents and H=0 kA/m. The trajectories can be analyzed in analogy to a previous work [19] of multimode generation regimes in single layer excitations. The trajectory $m(\psi)$ shown in Fig. 6 corresponds to what has been called a locked regime, where both modes are excited at the same frequency, as expected from eq. 11 in our coupled system. It can be seen in Fig. 6 that upon increasing current the mean value of $\psi$ approaches 90° as discussed in Fig. 3, and at the same time the mode asymmetry is reduced, i.e. the hybridization or mode mixing within the steady state mode increases. Note that previous work on such coupled system [10] already reported steady state modes that could not be defined as pure optical or pure acoustic modes. The results discussed here are in good agreement and actually give an explanation to the behavior and current dependence of the modes observed in ref 10.

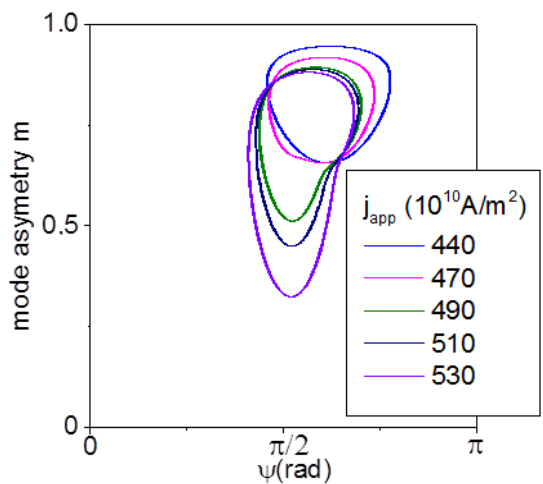

Fig. 6: Mode asymmetry at *H*=0 kA/m as a function of $\psi$ along a trajectory period at different currents.

## VI.B Influence of the mode mixing on the current dependence of the powers

Fig. 7a and 7b show the trajectory of power $p_2$ vs. power $p_1$ at $H=0$ kA/m and $H=3$ kA/m respectively at different currents. As it can be seen the dynamics is complex: the two modes are excited and their powers vary with time. Fig. 7c shows the mode asymmetry $<m>$ averaged over one period as a function of current at different fields. The range plotted for each value of field corresponds to the whole range of steady state excitations, from the critical current to the onset of chaotic excitations.

The first thing that should be pointed out is the difference in the mode asymmetry at the critical current when comparing low and high fields. At low fields (red symbols in Fig.7c) $<m>$ is close to 0.85 around the critical current, which means that the steady state mode is very strongly dominated by $b_1$. This can be seen also in the steady state power trajectory (black curve in Fig. 7a), which shows an almost negligible value of $p_2$ at the onset of steady state, for $j_{app}$=400 x $10^{10}$ A/m$^2$. Upon increasing the current density, $<m>$ is observed to decrease sharply (Fig. 7c), which corresponds to the steady state mode becoming a more hybridized mode. In other words, upon increasing current, $p_2$ increases more rapidly than $p_1$ (see Fig. 7d, red symbols). This shows that the energy supplied by the current is mainly used to enhance the amplitude $p_2$. In contrast to this behavior, at high fields (blue symbols in Fig. 7c) $<m>$ is already as low as 0.55 around the critical current. This corresponds to a relatively hybridized steady state mode with comparable contributions from $b_1$ and $b_2$ (see black curve in Fig. 7b). In this case an increase of the current does not translate into a strong change of $<m>$, which decreases but in a much smoother way (only a 19% reduction in the whole steady state range, while at low fields there is a 42% reduction, see Fig. 7c). This means that at high fields the impact of the current on $p_1$ and on $p_2$ is comparable (see Fig. 7d, blue symbols), a consequence of the initial hybridization of the steady state mode.

In conclusion, the current dependence of $p_1$, $p_2$ and $p_1/p_2$ (Fig. 7d) is different at low and high fields. While at low fields the energy supplied by the current goes mainly to the power of the non-dominant mode (in the case here $b_2$), at high fields it goes almost equally to $b_1$ and $b_2$. This is due to the different level of mixing of the steady state mode, which increases upon increasing field, as can be seen by the decrease of the mode asymmetry at the critical current upon increasing field (Fig. 7c).

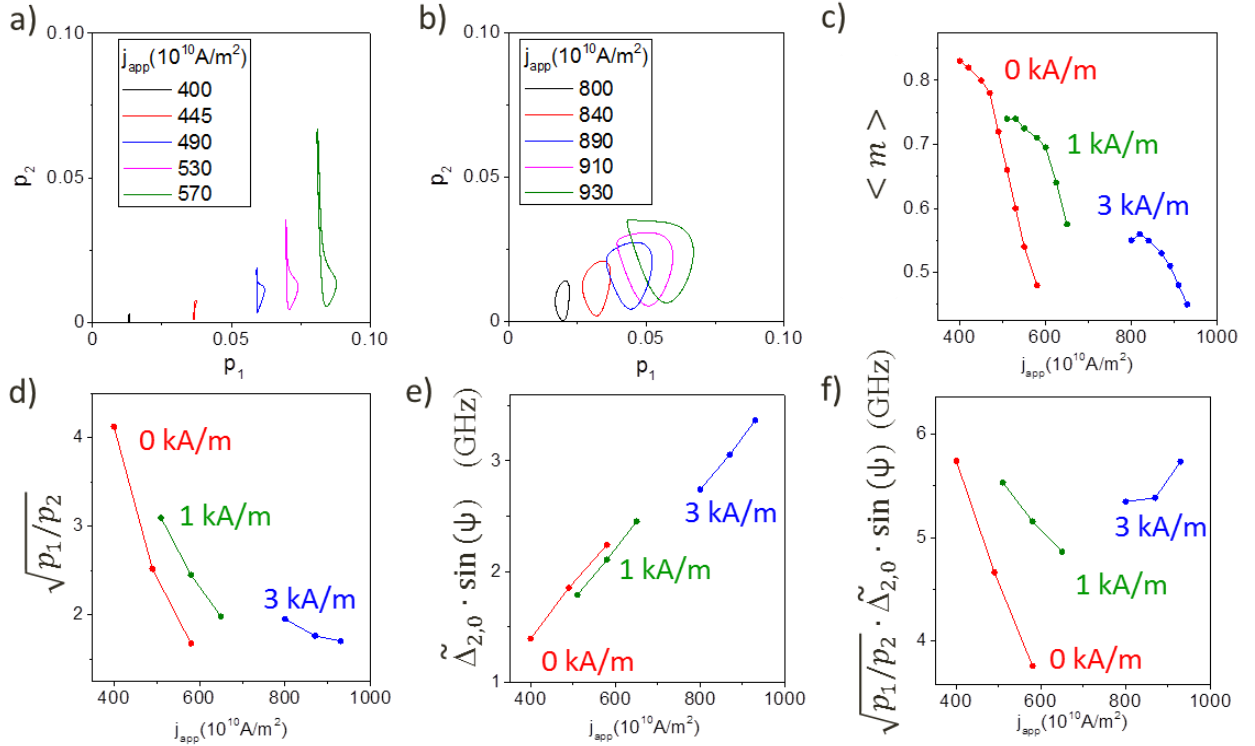

Fig.7: (a) $p_2$ vs $p_1$ at $H=0$ kA/m and (b) $H=3$ kA/m. (c) Current dependence of the mode asymmetry (averaged over a period), (d) $\sqrt{\frac{p_1}{p_2}}$, (e) $\widetilde{\Delta}_{2,0}\cdot\sin(\psi)$ and (f) $\sqrt{\frac{p_1}{p_2}}\cdot\widetilde{\Delta}_{2,0}\cdot\sin(\psi)$, at $H=0$ kA/m (black symbols), $H=1$ kA/m (red symbols) and $H=3$ kA/m (blue symbols).

## VI.C Consequence on the frequency-current dependence

The current dependence of $p_1$, $p_2$ and especially $p_1/p_2$ (see Fig. 7d) being very different at low and high fields has important consequences on the current dependence of the conservative and dissipative terms. Let us analyze here the change on the behavior of the most important terms of the phase equations.

First, we analyze the change of behavior at low and high fields of the most important contribution to the $<d\phi_2/dt>$ equation: i.e. the first order part of the last dissipative term of eq. 14,

$$\sqrt{\frac{p_1}{p_2}} \cdot \widetilde{\Delta}_{2,0} \cdot \sin(\psi) \tag{16}$$

Where $\widetilde{\Delta}_{2,0}$ is a real coefficient. Let us start by the power ratio prefactor. At low fields $\sqrt{p_1/p_2}$ (red symbols in Fig.7d) strongly depends on the current, being strongly reduced by 60% from its initial value within the steady state range. In contrast, due to the high hybridization of the steady state that has been explained in the previous section, this ratio does not change too much with current at high fields (12% reduction in the whole steady state range). On the other hand, Fig. 7e shows the current dependence of $\widetilde{\Delta}_{2,0} \cdot \sin(\psi)$, which shows an almost constant positive slope independently of the applied field. The same slope at low and high fields shows that the different current dependencies at low and high fields come mainly from the power prefactors and not from a change in the current dependence of the coefficients of the dissipative terms (for instance $\widetilde{\Delta}_{2,0}$), as already said. Finally Fig.7f shows the current dependence of the full term $\sqrt{\frac{p_1}{p_2}} \cdot \widetilde{\Delta}_{2,0} \cdot \sin(\psi)$ at different fields. As can be seen, it changes from a redshift to a blueshift current-dependence, as a consequence of the very different current dependencies of $\sqrt{p_1/p_2}$ at low and high fields.

A similar analysis can be applied to the main contribution to $<d\phi_1/dt>$, which is $\Omega_1$. $\Omega_1$ is a conservative term so its coefficients $\omega$, $N_1$, $N_2$, $T$ are independent of the current and therefore constant at a given field. At all fields, $\Omega_1$ gives rise to a redshift behavior (green symbols in Figs. 5a and 5c) given only by the increase of $p_1$ and $p_2$ with current. Due to the reasons previously explained, at high fields this increase of power upon increasing current is much smoother than at low fields. As a consequence, the frequency decreases due to $\Omega_1$, along the whole steady state range it is reduced from 1.6 GHz at low fields (Fig. 5a) to 0.8 GHz at high fields (Fig. 5c). On the other hand, the dissipative terms $\sqrt{\frac{p_2}{p_1}}[\widetilde{\Theta}_1 \sin(\phi_1 - \phi_2) + \widetilde{\Delta}_1 \sin(\psi)]$ give a blueshift contribution to the $<d\phi_1/dt>$ current dependence (see violet symbols in Fig. 5c). At a certain field, such a blueshift contribution becomes larger than the redshift contribution given by $\Omega_1$. This explains the transition from redshift to blueshift behavior in the $<d\phi_1/dt>$ equation.

## VII. CONCLUSIONS

From the analysis of the different terms of equation 9 and its comparison to the numerical simulations, it is possible now to truncate the general perturbed Hamiltonian equation 7. The simplified perturbed Hamiltonian equation can be written as follows:

$$\begin{pmatrix} \dot{b}_1 \\ \dot{b}_2 \\ \dot{b}_1^* \\ \dot{b}_2^* \end{pmatrix} = -\begin{pmatrix} i(\Omega_1 + \widetilde{\Psi}_1) + \widetilde{\Gamma}_1 & \widetilde{\Theta}_1 & \widetilde{\Xi}_1 & \widetilde{\Delta}_1 + i\vartheta_1 \\ \widetilde{\Theta}_2 & i(\Omega_2 + \widetilde{\Psi}_2) + \widetilde{\Gamma}_2 & \widetilde{\Delta}_2 + i\vartheta_2 & \widetilde{\Xi}_2 \\ \widetilde{\Xi}_1 & \widetilde{\Delta}_1 - i\vartheta_1 & -i(\Omega_1 + \widetilde{\Psi}_1^*) + \widetilde{\Gamma}_1 & \widetilde{\Theta}_1 \\ \widetilde{\Delta}_2 - i\vartheta_2 & \widetilde{\Xi}_2 & \widetilde{\Theta}_2 & -i(\Omega_2 + \widetilde{\Psi}_2^*) + \widetilde{\Gamma}_2 \end{pmatrix}\begin{pmatrix} b_1 \\ b_2 \\ b_1^* \\ b_2^* \end{pmatrix} \tag{18}$$

Where

$\widetilde{\Psi}_1 = Fp_2 e^{j2(\phi_1 + \phi_2)} + \Lambda_1$
$\widetilde{\Psi}_2 = Fp_1 e^{j2(\phi_1 + \phi_2)} + \Lambda_2$

And where $\widetilde{\Gamma}_i$, $\widetilde{\Theta}_i$, $\widetilde{\Delta}_i$, $\vartheta_i$ are real functions of powers such as $\chi_i = \chi_{i,0} + \chi_{i,1}|b_1|^2 + \chi_{i,2}|b_2|^2$, where the coefficients $\chi_{i,i}$ are real values depending on material parameters, injected current and damping constant.

Note that all off-diagonal terms in eq. 18 have a dissipative origin. The *antidiagonal* terms ($\widetilde{\Delta}_i + i\vartheta_i$) are those with strongest impact in the dynamics.

To our knowledge, most previous works related to mode interaction have taken into account either only the conservative terms or the conservative terms and the diagonal dissipative terms ($\widetilde{\Gamma}_i$), usually considered to be the effective damping rate of the *i*th mode power [20]. Some studies include one off-diagonal dissipative term (namely $\widetilde{\Theta}_{i,0}$) in the $b_1$ and $b_2$ equations to describe linear interaction or coupling between modes [21]. The other off-diagonal dissipative terms ($\widetilde{\Xi}_i$, $\widetilde{\Delta}_i$, $\vartheta_i$) are not intuitively evident and, to our knowledge, none of them have been previously considered, neither studying multimode generation regimes in single layers excitations [21] nor phase locking between different oscillators [20]. As shown here, these terms can be very important in coupled systems. Actually, the term $\widetilde{\Delta}_i$ happens to be the major contribution to the phase equation of the non-excited mode in this self-polarized system. Further analysis shows that these terms are also important to correctly predict the critical current from the analytical model [16]. For the externally polarized system the constant phase relation is different ($\psi_{ex}$ instead of $\psi$), and as a consequence the major contribution to the phase equation of the non-excited mode in the steady state regime for that system ($\propto \sqrt{\frac{p_j}{p_i}} \sin(\psi_{ex})$) is coming from $\widetilde{\Theta}_i$. From these results it is clearly concluded that the perturbed Hamiltonian equations of coupled systems contains off-diagonal dissipative terms such as $\widetilde{\Xi}_i$, $\widetilde{\Theta}_i$ and $\widetilde{\Delta}_i$ and those cannot be assumed to have a negligible impact.

To summarize, in this work the non-linear auto-oscillator theory of a single layer magnetization dynamics has been adapted to a simple two layer coupled system. An analytical model has been proposed and validated through numerical simulations. First of all, the basis transformation required to diagonalize the quadratic part of the Hamiltonian has been proposed and applied to the dissipative part of the LLGS equation, in order to define a general model. Secondly, the higher order terms of the model have been analyzed and tested against numerical simulations to better understand the dynamics of this system. Many terms should be included in the model to obtain accurate frequency values. A simplified model of analytical oscillator equations that describe the non-linear dynamics was then proposed and validated through numerical simulations. The main findings are:

(i) In contrast to the single layer model, the phase equation of the coupled system has a contribution coming from the dissipative part of the LLGS equation. It is shown that this is a mandatory contribution to describe well the frequency of the steady state and the most basic features of coupled dynamics.

(ii) The conservative term coming from the S-term of the Hamiltonian does not play a role in the dynamics of the self-polarized SyF, unlike in most other coupled systems.

(iii) In order to describe the STT steady state excitations, both linear eigenmodes contribute and the characteristic relation between the phases of the eigenmodes is such that the *sum* of the phases remains constant, and not the phase difference as it is the case in most phase locked systems.

(iv) A specific feature of coupled dynamics is addressed: the redshift to blueshift transition observed in the frequency-current dependence upon increasing the applied field [11]. It is found that it comes from the competition between conservative and dissipative terms, where the non-trivial dependence of these terms on the amplitudes and phases plays a crucial role. It is found that the blueshift regime can only occur when the power ratio $p_1/p_2$ has a weak dependence with current. This happens only in a region of field where the two linear eigenmodes contribute equally to the steady state mode (i.e. high mode hybridization).

(v) Finally, a general perturbed Hamiltonian equation was proposed for this coupled system, giving the tools to study other kind of coupled systems (it has been tested for instance in externally polarized coupled system) and to address other questions of coupled dynamics.

This analytical model opens perspectives to better control and understand non-linear phenomena and it will be important to further predict the non-linear parameters as a function of coupling strength, and to exploit the coupling mechanisms to optimize the microwave performances for applications.

**Appendix 1:**

Canonical transformation that connects the cartesian magnetization components $m_{x,y,z}(t)$ to the complex variables $a(t)$ and $a^*(t)$:

$$a_1 = \beta^{-1/4} \frac{m_{Y1} - jm_{Z1}}{\sqrt{2(1+m_{X1})}} \quad \text{and} \quad a_2 = \beta^{1/4} \frac{m_{Y2} - jm_{Z2}}{\sqrt{2(1+m_{X2})}}$$

$$a_1^* = \beta^{-1/4} \frac{m_{Y1} + jm_{Z1}}{\sqrt{2(1+m_{X1})}} \qquad a_2^* = \beta^{1/4} \frac{m_{Y2} + jm_{Z2}}{\sqrt{2(1+m_{X2})}}$$

where $\beta = \frac{t_2}{t_1} \frac{M_{s2}}{M_{s1}}$ is the asymmetry factor.

The reversed transformation is given by:
$$m_{x1} = -m_{X1} = -(1 - \beta^{1/2} 2 a_1 a_1^*)$$
$$m_{y1} = -m_{Y1} = -\beta^{1/4}(a_1 + a_1^*)\sqrt{1 - \beta^{1/2} a_1 a_1^*}$$
$$m_{z1} = m_{Z1} = j\beta^{1/4}(a_1 - a_1^*)\sqrt{1 - \beta^{1/2} a_1 a_1^*}$$
$$m_{x2} = m_{X2} = 1 - \beta^{-1/2} 2 a_2 a_2^*$$
$$m_{y2} = m_{Y2} = \beta^{-1/4}(a_2 + a_2^*)\sqrt{1 - \beta^{-1/2} a_2 a_2^*}$$
$$m_{z2} = m_{Z2} = j\beta^{-1/4}(a_2 - a_2^*)\sqrt{1 - \beta^{-1/2} a_2 a_2^*}$$